\documentstyle[twoside,fleqn,espcrc2,psfig,epsf]{article}

\def\A{\kern +0.6ex\lower 0.42ex\hbox{$\scriptstyle\bf\iota$}\kern -1.20ex a}

\newcommand{\be}{\begin{equation}}
\newcommand{\ee}{\end{equation}}
\newcommand{\ba}{\begin{eqnarray}}
\newcommand{\ea}{\end{eqnarray}}

\newcommand{\AmS}{{\protect\the\textfont2
  A\kern-.1667em\lower.5ex\hbox{M}\kern-.125emS}}

\title{\textbf{The Tau Lepton: Particle Physics in a Nutshell}}
\author{Johann K\"uhn
\address{Institut f\"{u}r Theoretische Teilchenphysik,\\
Universit\"{a}t Karlsruhe,
D--76128 Karlsruhe, Germany}}

\begin{document}
\begin{abstract}
\end{abstract}
\maketitle

{\bf Outline}\\
1. Introduction\\
2. Weak Couplings\\
\hspace*{0.5cm}2.1 Charged current interactions\\
\hspace*{0.5cm}2.2 Neutral current interactions\\
\hspace*{0.5cm}2.3 Electric and magnetic dipole moments\\
3. Inclusive Decays, Perturbative QCD and Sum Rules\\
\hspace*{0.5cm}3.1 Inclusive decays and the strong coupling\\
\hspace*{0.5cm} constant\\
\hspace*{0.5cm}3.2 Cabibbo suppressed decays and the\\ 
\hspace*{0.5cm}Strange Quark Mass\\
4. Exclusive Decays\\
\hspace*{0.5cm}4.1 Form factors and structure functions\\
\hspace*{0.5cm}4.2 Chiral dynamics\\
\hspace*{0.5cm}4.3 Resonances\\
\hspace*{0.5cm}4.4 Isospin and CVC\\
\hspace*{0.5cm}4.5 Hadronic vacuum polarization from \\
\hspace*{0.5cm}$\tau$ decays\\
5. Beyond the Standard Model\\
\hspace*{0.5cm}5.1 CP violation in hadronic $\tau$ decays\\
\hspace*{0.5cm}5.2 ``Forbidden'' decays.

\section {Introduction}

More than 20 years after its discovery, the study of the
tau lepton remains a fascinating field, encompassing particle physics
in its full variety, from strong to electromagnetic and
weak interactions, from resonance physics at long distances to 
perturbative short distance physics.  The Standard Model in its
large variety of phenomena is of relevance as well as subtle tests
of its validity and searches for physics beyond this well-explored
framework.  As a member of the third family with its large mass, more
than three thousand times more massive than the electron, it could be
particularly sensitive to new interactions related to the Higgs
mechanism.

Specifically, as indicated in Fig.1, the production process in
electron-positron collisions allows to explore the lepton
couplings to photon and the Z boson, its charge, magnetic and
electric dipole moment, the vector and axial couplings of both
electron and tau lepton, thus providing one of the most precise
determinations of the weak mixing angle.  Its weak decay
gives access to its isospin partner
$\nu_{\tau}$,
providing stringent limits on its mass
$m_{\nu}$
and its helicity
$h_{\nu}$.
Universality of charged current interactions can be tested
in a variety of ways which are sensitive to different scenarios
of physics beyond the Standard Model.

The hadronic decay rate and, more technically, moments of 
the spectral function as calculated in perturbative QCD lead to
one of the key measurements of
$\alpha_s$, remarkable
in its precision as well as its theoretical rigor, and
the Cabibbo suppressed transition might well allow for an
accurate determination of the strange quark mass.

Resonances, and finally
$\pi$, K, $\eta$, and $\eta\prime$
are the hadronic decay products, and any theoretical
description of
$\tau$
decays should in the end also aim at an improved understanding
of this final step in the decay process.  At low momentum
transfer one may invoke the technology of chiral Lagrangians,
for larger masses of the hadronic system vector dominance
leads to interesting phenomenological constraints.  With
increasing multiplicity of the hadronic state the transition
matrix elements of the hadronic current are governed by form
factors of increasing complexity.  Bilinear combinations of these
form factors denoted ''structure functions'' determine the angular
distributions of the decay products.  In turn, through an appropriate
analysis of angular distributions it is possible to reconstruct
the structure functions and to some extent
form factors.

Isospin relations put important constraints on these analyses and
allow to interrelate hadronic form factors measured in tau decays
with those measured in electron-positron annihilation.

Obviously, there remains the quest for the unknown, the truly
surprising.  CP violation in tau decays, extremely suppressed
in the framework of the Standard Model, is one of these options,
transitions between the tau and the muon or the electron, or even
lepton number violating decays are other exciting possibilities.
This overview also sets the stage for the present talk: 
Sect.\ref{S2} will be concerned with the tau as a tool to explore weak
interactions, charged and neutral current phenomenology, in
particular the test for universality, the determination of the
neutrino helicity, and searches for anomalous couplings.  Sect.\ref{S3}
will be concerned with perturbative QCD, in particular with recent
results on the beta function, the anomalous dimension and their
impact on current analyses of 
$\alpha_{s}$.

\begin{figure*}
  \begin{center}
    \leavevmode
    \epsfxsize=10.cm
    \epsffile[80 60 550 380]{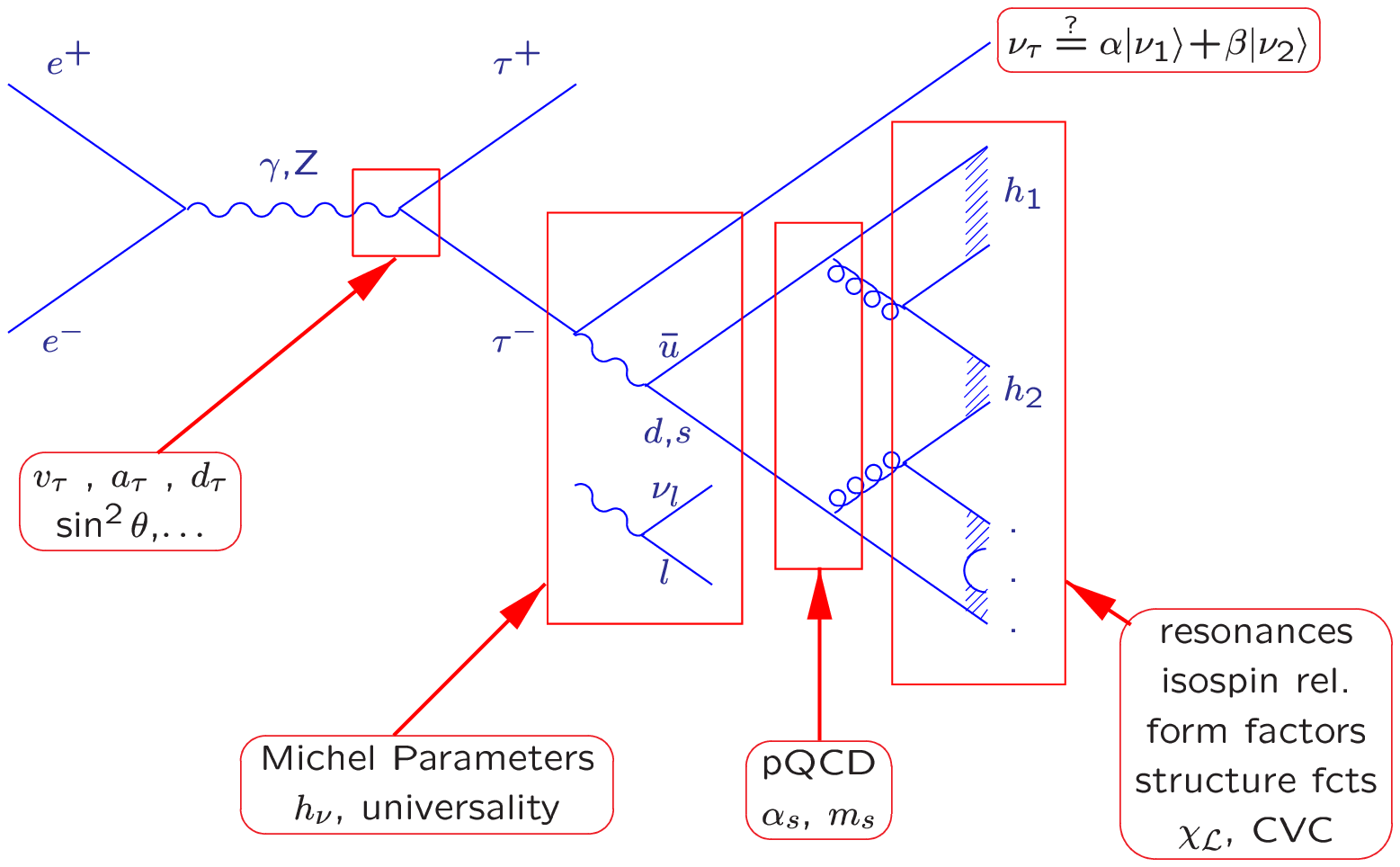}
    \hfill
    \parbox{14.cm}{
    \caption[]{\label{F1}\sloppy
Selected physics topics which can be studied in $\tau$-lepton
production and decay.}}
  \end{center}
\end{figure*}

Recent higher-order calculations of the interdependence between
the strange quark mass and the decay rate into Cabibbo
suppressed channels are another important 
topic of Sect.\ref{S3}.  Predictions for exclusive decays will
be covered in Sect.\ref{S4}.  This includes the technique of structure
functions, predictions for form factors based on chiral dynamics,
the inclusion of resonances, and constraints from CVC and
isospin relations.  Tau decays in combination with isospin have 
been used to measure the pion form factor which in turn is an
important ingredient for the calculation of the anomalous magnetic
moment of the muon and the electromagnetic coupling at the
scale
$M_{Z}$.
The validity of this approach and its limitations will also be
addressed in Sect.\ref{S4}.
A  selection from the many speculations on physics
beyond the Standard Model will be presented in Sect.\ref{S5}. This
includes tests for CP violation and a few new suggestions for lepton
number violating decays.

The properties of the tau neutrino, in particular its mass and mixing,
have received considerable attention after the observation of mixing
in atmospheric neutrino studies.  This topical subject will be
discussed in a different section of this meeting and is therefore
not included in the present review.

\section {\label {S2} Weak Couplings}

\subsection{\label{Ss21} Charged Current Interactions}

\subsubsection{\label {Sss21} Lepton Universality in Tau Decays}
Universality of weak interactions is one of the cornerstones
of the present theoretical framework.  Deviations could 
arise from additional gauge interactions which are sensitive
to lepton species, thus providing a clue to the origin of the
triplicate nature of the fermion spectrum, one of the mysteries
of the present theory. Alternatively one might attribute a
deviation from universality to mass dependent
interactions, mediated for example by Higgs exchange in 
fairly straightforward extensions of the Standard Model.\\

The different tests of universality as discussed below are
sensitive to different new phenomena.  They should therefore
be pursued as important measurements in their own right, and 
not simply be judged and compared on the basis of one
''figure of merit''.\\

{\it i}) $\tau \to \nu \pi$\\

The classical test of lepton universality dating back to the
``pre-tau-era'' is based on the comparison between the
decay rates of the pion into electron and muon respectively.
In lowest order this ratio is simply given by the electron and
muon masses, a simple first year's textbook calculation. 
A precise prediction must include radiative corrections (Fig.2)
due to virtual photon exchange and real emission 
\ba
&& R_{e/\mu} = \frac {\Gamma(\pi \to e \nu)}{\Gamma(\pi \to \mu \nu)}
\nonumber \\
&&= \frac {m_e^2}{m_\mu^2} \left (\frac {m_\pi^2 -
m_e^2}{m_\pi^2-m_\mu^2} \right )^2
\left (1+\delta R_{e/\mu} \right ).
\ea

\begin{figure*}
\begin{center}
\mbox{}\hspace{.5cm}\epsfxsize=7.cm\epsffile[180 340 420 470]{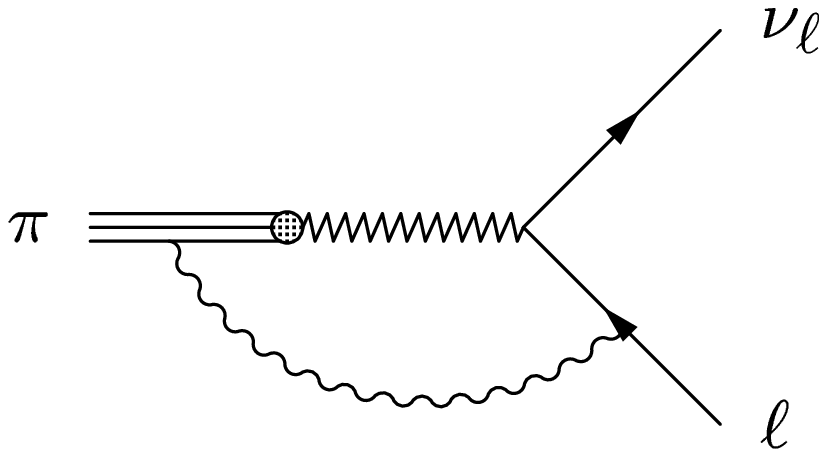}
\parbox[b]{10cm}{
\parbox[b]{.8cm}{+\vspace{1.5cm}}
\epsfxsize=7.cm\epsffile[180 340 420 470]{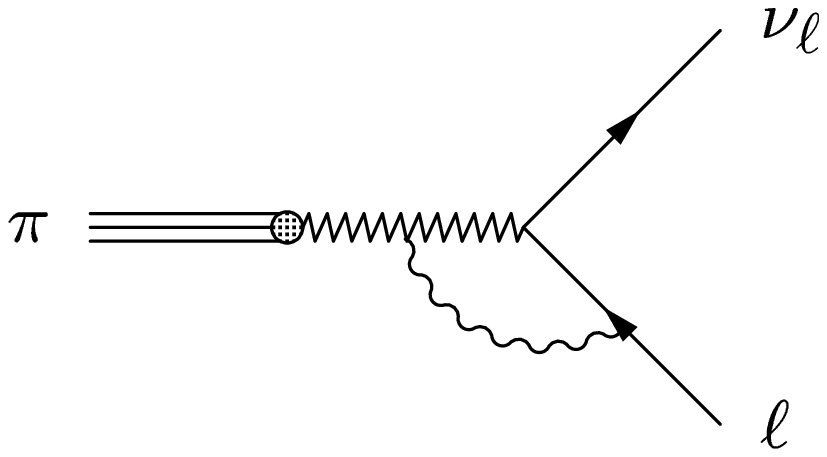}\parbox[b]{2.cm}{+\,\ldots
\vspace{1.5cm}}}\\[2mm]
\quad {{\bf + real emission}}
    \caption[]{\label{F2}\sloppy
  Radiative corrections affecting the pion decay.  
 }
\end{center}
\end{figure*}

Short distance corrections cancel in 
$R_{e/\mu}$ since it is the same effective Hamiltonian
which governs both numerator and dominator. The
dominant $m_{\ell}$ dependent correction, 
\be
R_{e/\mu} = \frac {3\alpha}{\pi} \ln \frac {m_e}{m_\mu} + ...
= -3.72 \%
\ee
comes remarkably close to a more refined prediction
including renormalization group improvement plus an
estimate of the dominant logarithms from hadronic effects
\cite{MS1}
\be
\delta R_{e/\mu}=-(3.76 \pm 0.04) \%.
\ee
A more refined treatment \cite{DF} including a detailed
modeling of hadronic form factors   leads to
\be
\delta R_{e/\mu}=-(3.74 \pm 0.01) \% ,
\ee
well consistent with the earlier result.
The reduced theoretical uncertainty can again 
be traced to the long distance aspect
of hadronic interactions.

The final prediction and the measurement are well consistent:
\ba
&&R^{\rm theory}_{e/\mu}=(1.2356 \pm 0.0001)\cdot 10^{-4},
\\
&&R^{\rm exp}_{e/\mu}=(1.230 \pm 0.004)\cdot 10^{-4}.
\ea
Universality in this reaction is thus verified at a level of 
$3\cdot 10^{-3}$
and further progress is possible by improving the
experimental precision.

The small uncertainty of the theoretical prediction
is a natural consequence of the low momenta probed
in this calculation, with $m_\pi$ far below $m_\rho$, the
typical scale relevant for hadronic form factors. 
The situation is drastically different for tau decays.
In lowest order the ratio 
\ba
&&R_{\tau/\pi} = \frac {\Gamma(\tau \to \pi \nu)}{\Gamma(\pi \to \mu \nu)}
\nonumber \\
&&=  \frac {m_\tau^3}{2m_\pi m_\mu^2} \left (
\frac {1-m_\pi^2/m_\tau^2}{1-m_\mu^2/m_\pi^2} \right )^2
\left (1+\delta R_{\tau/\pi} \right )
\ea
 is again fixed by the
masses of the leptons and the pion.  The radiative
corrections collected in $\delta R_{\tau/\pi}$ 
can again be predicted
in leading logarithmic approximation \cite{MS2}
\be
1+\delta R_{\tau/\pi} = 
\frac {1+2\frac {\alpha}{\pi}\ln m_Z/m_\tau}
{1+\frac {3}{2} \frac {\alpha}{\pi}\ln m_Z/m_\pi 
+\frac {1}{2} \frac {\alpha}{\pi}\ln m_Z/m_\rho}. 
\label{e1}
\ee
The short distance piece as given by the
$\ln m_{Z}$  terms cancels again in the ratio, the remainder depends
critically on the guess for the low energy cutoff.  The 
choice as given in Eq.(\ref{e1}) leads to 
\be
\delta R_{\tau/\pi}= -0.96 \%
\label{e2}
\ee
with an uncertainty estimated in \cite{MS2} to be $\pm 0.5 \%$.
A detailed explicit calculation based on a careful separation
between long and short distance contributions predicts \cite{DF}
\be
\delta R_{\tau/\pi} = \left (0.16^{~~+0.09}_{~~-0.14} \right ) \%
\label{e21}
\ee
The difference between these two results is entirely due to the more detailed
treatment of long distance phenomena.  The important lesson to be drawn
from this comparison is that estimates of long distance effects for
exclusive channels may well fail at the level of $1 \%$.  This uncertainty
of around $1 \%$ might also set the scale for predictions of 
$\Gamma(\tau \to \pi^- \pi^0 \nu)$
based on CVC and isospin symmetry.

The prediction for decays into $\pi$ and similarly into $K$
is well consistent with the experimental result
\be
B_{\pi+K} = (11.79 \pm 0.12)\% .
\ee
The lifetime $\tau_\tau = 290.8 \pm 0.6~{\rm fs}$  
which has to be used in this comparison as additional input is
derived from the direct measurement
$\tau_\tau = 290.5 \pm 1.0~{\rm fs}$
and the indirect result
of $291.0 \pm 0.7~{\rm fs}$ 
which is obtained from the leptonic
branching ratios $B_e = 17.81 \pm 0.06$\% and
$B_\mu = 17.36 \pm 0.06$\%. 
The test of universality, ${\rm Exp/Th} = 1.013 \pm 0.010$, is thus only
a factor three less precise than the universality test in pion decay.\

It is interesting to see that experiments with their accuracy of 
one percent are already able to discriminate 
between the theoretical predictions Eq.(\ref{e2}) and Eq.(\ref{e21}). 
The former, based on leading logarithm 
considerations only, disagrees by more than
two standard deviations, the latter is  consistent with the
measurements.  Future, improved measurements of 
$\tau \to \nu \pi$  will therefore
be critical tests of our understanding of radiative corrections, with
important implications for the validity of isospin symmetry and CVC
in the prediction of $\tau \to \nu \pi^- \pi^0$
from $e^{+}e^{-}$ data discussed below (Sect. 4).

Although the present result for the mode 
$\tau \to \nu K$  with a branching
ratio $0.71 \pm 0.05$\% is not yet competitive it should be
emphasized that this channel is particularly interesting.
It connects quarks and leptons of the second and third generation,
respectively.  It is particularly sensitive to exotic mass dependent
effects such as charged Higgs exchange and its rate is
predicted with an accuracy better than one percent.\\

{\it ii})~~Leptonic decays\\

The rates of two of the three purely leptonic decays can be
calculated without theoretical ambiguity if the third one has
been measured.  Conventionally one chooses the muon decay rate
for normalization.  The electromagnetic corrections are finite 
and given separately in the implicit definition of $G_{F}$
\ba
&&\Gamma_{\mu} = \frac {G_F^2 m_\mu^5}{192 \pi^3} \left (
1+\frac {3m_\mu^2}{5m_W^2} \right ) 
f(\frac {m_\mu^2}{m_W^2})
\left (1+\delta_{\rm QED} \right ),
\nonumber \\
&&
f(x) = 1-8x+8x^3-x^4-12x^2\ln x
\\
&&
\delta_{\rm QED} = \left (\frac {\alpha \left (m_\mu \right)}{\pi}\right ) 
\left (\frac {25}{8}-\frac
{\pi^2}{2} \right ) + \left (\frac {\alpha}{\pi} \right )^2 6.743.
\nonumber
\ea
The ${\cal O}(\alpha)$ term in $\delta_{QED}$  
has been calculated nearly forty years
ago \cite{muon1}, the second term of order $\alpha^{2}$ was evaluated only
recently \cite{Timo}.

Two independent tests of universality are thus accessible: the
comparison of electronic and muonic decay rates of the $\tau$,
and the comparison between one of the tau decay rates and 
$\Gamma_\mu$.

Electron-muon universality is best tested in the ratio 
$B_{\mu}/B_e$. The experimental result, expressed in terms of the ratio 
of couplings \cite{Passaleva} 
\be
\frac {g_\mu}{g_e} = 1.0015 \pm 0.0025.
\ee
comes close in precision to the one 
achieved in pion decay $(1.0023 \pm 0.0016)$.
Higgs exchange amplitudes could in particular be enhanced in 
$\tau \to \mu \nu \nu$ and thus lead to observable differences
in $B_e/B_\mu$ (Fig.3). In pion decays they would be 
practically absent.

\begin{figure}
\caption[]{\label{F3}\sloppy Amplitudes mediated by Higgs exchange.
      }
 \begin{center}
    \leavevmode
    \epsfxsize=10.cm
\hspace{.5cm}\epsfxsize=7.cm\epsffile[135 335 420 470]{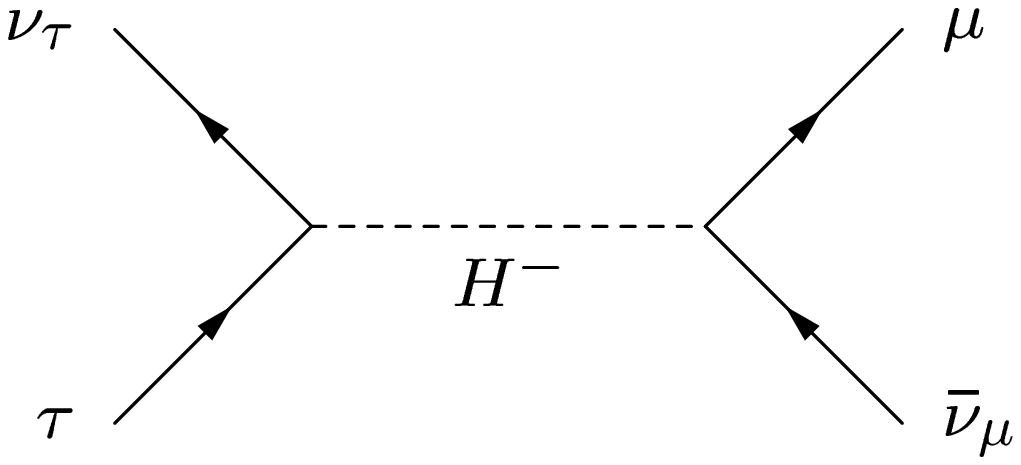}\quad
\epsfxsize=7.cm\epsffile[180 335 420 470]{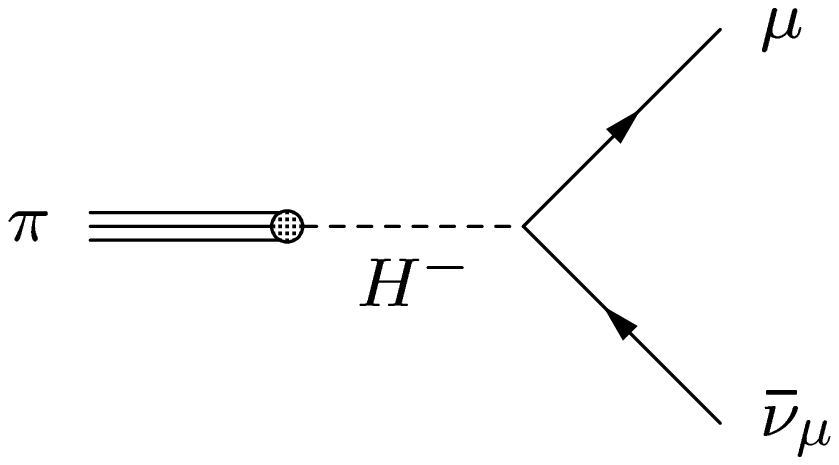}
    \hfill
  \end{center}
\end{figure}

The relative
reduction of the rate in the two Higgs doublet model 
$-2m_\mu^2  ({\rm tan}\beta /m_H )^2$
leads to
an interesting limit on $m_H/{\rm tan}\beta$ 
in the range around or above $1$ GeV \cite{Stahl,Stahltp},
nearly comparable in strength to those from 
$b \to s \gamma$, $B \to \tau \nu$, and $Z \to b \bar b$.
Additionally, even stronger limits have been deduced from the decay
spectra, the Michel parameters, to be discussed below.

The ratio $B_{\mu}/B_{e}$ thus leads to a test of
electron-muon universality. The comparison between the leptonic
decay rate of the tau and the muon is in contrast sensitive to
new physics phenomena connected specifically 
with the tau lepton.
One example \cite{Guth} is based on enhanced Higgs boson induced vertex
corrections, which leads to a relative reduction of the rate
\be
\frac {\delta \Gamma}{\Gamma} = \frac {G_F m_\tau^2}{16 \pi^2}\;
2\sqrt{2}\; {\rm tan}^2\beta \; {\cal F}\left ( m_{H_i},\alpha_{\rm mix}
\right ),
\ee
where ${\rm tan}\beta$ is given by the ratio of the two vacuum
expectation values and $\alpha_{\rm mix}$ parameterizes the 
mixing in the Higgs sector.  Depending on the precise
values of these parameters the reduction of the rate might reach
a few per mille.  This is indicated in Fig.4 for the specific
choice $m_{H_{1}} = m_{H_{2}} = m_{H_{3}} = m_{H}$ labeling
the curves, ${\rm tan}\beta = 70$, and the charged Higgs mass 
varying between $50$ and $1000$ GeV. 

\begin{figure}
\caption[]{\label{F4}\sloppy 
The ratio $1+\delta\Gamma/\Gamma$ versus the mass
of the charged  Higgs boson for
$m_{H_1}=m_{H_2}=m_{H_4}=$
$40-400$ GeV and 
${\rm tan}\beta=70$ \cite{Guth}.}
  \begin{center}
    \leavevmode
    \epsfxsize=7cm
\epsffile{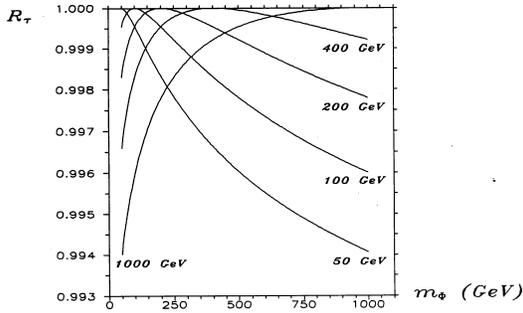}
    \hfill
  \end{center}
\end{figure}

 Larger mass splittings
typically lead to larger deviations. Present experiments
with their sensitivity of a few permille \cite{Passaleva}
\be
\frac {g_\tau}{g_\mu} = 1.0001 \pm 0.0026
\ee
start to approach this interesting region.

\subsubsection{ \label{Sss212}
Lepton Spectra, Michel Parameters, and Neutrino Helicity}

{\it i}) Leptonic decays
\vspace*{0.5cm}

Much of the techniques of
measuring and analyzing the lepton spectrum in tau decays has been 
derived from the corresponding muon decay experiments.  Starting
from the local four fermion interaction
\be
{\cal M}= \frac {4G_F}{\sqrt{2}}\sum g_i \left (\bar l_{\rm L,R}
\Gamma^N \nu_l \right ) \left (\bar \nu_\tau
\Gamma_N \tau_{\rm L,R} \right )
\label{e4}
\ee
and exploiting either the ``natural'' polarization of $\tau$'s
from $Z$ decays or the ``induced'' polarization from $\tau^{+}\tau^{-}$
decay correlations the four parameters 
$\rho,\eta,\xi$ and $\delta$
are determined consistent
with the expectations from V-A interaction $3/4$, $0$, $1$ and $3/4$.  The
parameter $\eta$ in particular is sensitive to righthanded
couplings; its measurement can be used to set interesting bounds
on amplitudes induced through charged Higgs exchange  
\cite{Stahl,Stahltp} quite comparable to those derived from the rate.

Recently the local ansatz Eq.(\ref{e4})
has been questioned, and a generalization
including momentum dependent vertices has been
introduced \cite{Taylor}.  Starting from a coupling of the form
\be
{\cal A} = \frac {g}{\sqrt{2}}\bar \tau \left [
\gamma_\mu + \frac {i}{2m_\tau}\sigma_{\mu \nu}
Q^{\nu} \left ( \kappa - i\tilde \kappa \gamma _5 \right ) \right ]
\nu _{\rm L} W^\mu
\ee
with $Q = P_\tau - P_\nu$ , one 
obtains modifications of the distributions in leptonic
and semileptonic decays which are not covered by the Michel parameters.
A limit on the anomalous coupling $\kappa$ obtained this way is easily 
converted into a limit on the compositeness scale $\Lambda =m_\tau/\kappa$,
and experiments are getting close to interesting bounds in the
range $\Lambda \sim 100$ GeV.  
This topic is closely related to the discussion
of nonlocal neutral current interactions in Sect.2B.

\vspace*{1cm}
{\it ii}) Semileptonic decays and neutrino helicity
\vspace*{0.5cm}

\begin{figure}
 \caption[]{\label{F5}\sloppy
Three pion mode as a tool for \\the measurement of the neutrino helicity.}
  \begin{center}
    \leavevmode
    \epsfxsize=7cm
\epsffile[70 385 310 450]{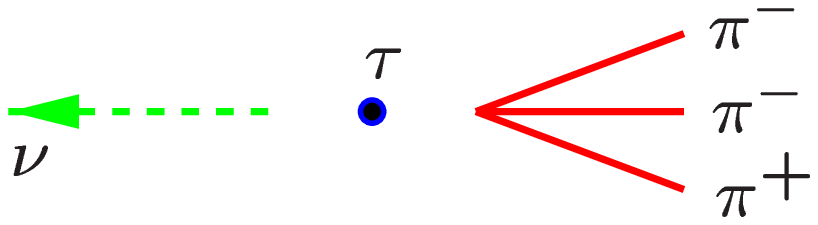}
    \hfill
  \end{center}
\end{figure}

Semileptonic decays offer two distinctly different methods to 
determine the relative magnitude of vector vs. axial amplitude in the
$\tau \nu W$ coupling.  Given a nonvanishing tau polarisation, the angular
distribution of hadrons relative to the direction of the polarization
is a direct measure of 
\be
h_\nu = \frac {2g_Vg_A}{g_v^2+g_A^2},
\ee
with  $h_\nu = -1$ in the V-A theory. For the pion decay 
for example one obtains, for example:
\be
dN = \frac {1}{2} \left (1 -h_\nu \cos \theta \right ) {\rm d} cos
\theta .
\label{sp}
\ee
For more complicated multi-meson final states the coefficient
accompanying $h_\nu$ is smaller than one, if the hadronic final
state is summed.  Full analyzing power is recovered by looking into
suitable angular distributions as discussed below.  An alternative
determination \cite{KW} of $h_\nu$, applicable in particular
for unpolarized taus,
is provided by decays into three (or more) mesons.  In this case
one can discriminate between hadronic states with helicity $-1$, zero
or $+1$ and thus infer the neutrino helicity from the
analysis of parity violating (!) angular 
distributions of the mesons (Fig.\ref{F5}).

As mentioned above, the analyzing power of two- or three-meson
states in a spin one configuration is reduced by a factor 
$(m_\tau^2-2 Q^2)/(m_\tau^2+2Q^2)$
which is typically around $0.5$ for two pions 
($Q^2 \approx m_\rho^2$)  and close to zero 
for three pions $Q^2 \approx m_{a_1}^2$.  
Full analyzing power of one, equal to the
single pion decay, is recovered if the information on all
meson momenta is retained.  Starting from the decay amplitude
\be
{\cal M} = \frac {G_F}{\sqrt{2}} \bar u(\nu) \gamma^\alpha (1-\gamma_5)
u(\tau) J_\alpha
\ee
with $J_\alpha = \langle {\rm had} | V-A| 0 \rangle$
one obtains
\be
{\rm d}\Gamma =\frac {G_F^2}{2m} \left ( \omega 
- {\bf H}{\bf S} \right ) {\rm d}PS.
\ee

For the single pion decay ${\bf H}/\omega = h_\nu {\bf n}_\pi$. 
In general  ${\bf H}$ and $\omega$ are constructed
from a bilinear combination of the hadronic current dependent
on all meson momenta.  As shown in \cite{KW,K}
\be
|{\bf H}/\omega | = 1.
\ee
This observation is also one of the important
ingredients of the Monte Carlo program TAUOLA \cite{Tauola},
 which is currently used to simulate tau decays.  
It has furthermore been used \cite{Da} 
for a simplified
analysis of the tau polarization in terms of one ``optimal variable''.
The complete kinematic information is not immediately available in
these events since the momenta of the two neutrinos from the
decay of $\tau^+$ and $\tau^-$  respectively are not determined
and one is left with a twofold ambiguity, 
even in double hadronic events.
This leads to 
a reduction of the
analyzing power for two and three meson decays.  However, as shown in 
\cite{K1}
the measurement of the tracks close to the production point 
with vertex detectors allows to resolve the ambiguity. Specifically,
it is the direction of the vector ${\bf d}_{\rm min}$
which characterizes the distance between the tracks of
$\pi^+$ and $\pi^-$ and which provides the missing piece of information
for events where both taus decay into one pion each. 
The situation for more complicated modes is discussed in
\cite{KM2}

Only recently, with significant advance in the vertex detectors
this technique has been applied, providing a factor two improvement
compared to the result without reconstruction of the full
kinematics \cite{ALEPH}.

\subsection{\label {Ss2.2} Neutral Current Couplings}

$Z$ decays into tau pairs provide an extremely powerful tool
for the analysis of neutral current couplings.  This has been
made possible by the large event rates collected by the four
LEP detectors and the inclusion of most of the tau decay modes
in the analysis.  At present these measurements compete with
those from the SLD experiment based on longitudinally polarized
beams, thus providing an important test of lepton universality
in the neutral current sector and an accurate independent
determination of the weak mixing angle.  A first important
test is obtained from a comparison of the rate \cite{rate}
\ba
&&R_e=20.757 \pm 0.056; 
\nonumber \\
&&R_\mu = 20.783 \pm 0.037; 
\nonumber \\
&&R_\tau = 20.823 \pm 0.056,
\ea
which confirms lepton universality of neutral currents at a level
comparable to the charged current result.

The $\tau$ polarization as function of the production angle $\theta$
depends on both electron and tau couplings:
\be
P_\tau (\cos \theta) = -\frac {A_\tau(1+\cos ^2 \theta ) + 2 A_e \cos \theta}
{(1+\cos ^2 \theta ) + 2 A_\tau A_e \cos \theta }
\ee
and the asymmetry coefficients
\be
A_l = \frac {2(1-4\sin^2 \theta_W)}{1+(1-4\sin^2 \theta_W)}
\ee
are extremely sensitive to the effective weak
mixing angle $\sin^2 \theta_W$:
\be
\delta A_l \approx 8 \delta \sin^2 \theta_W.
\ee


\begin{table}
\caption[]{\label {T1}Preliminary $A_e$ and $A_\tau$ results from ALEPH 
Collaboration
with statistical and systematic
uncertainties for the $1990-1995$ 
data based on the $\tau$ direction method. From \cite{Alemanytp}.}
\renewcommand{\arraystretch}{1.3}
\begin{center}
{\small
\begin{tabular}{|l|l|l|}
\hline\hline
Channel & $A_\tau(\%)$ & $A_e(\%)$\\
\hline
 hadron & $15.49 \pm 1.01 \pm 0.66 $ & $ 17.36 \pm 1.35 \pm 0.13$\\
rho & $13.71 \pm 0.79 \pm 0.57$&
 $15.04 \pm 1.06 \pm 0.078$\\
$a_1(3h)$ &$15.01 \pm 1.55 \pm 1.30$& $15.78 \pm 2.07 \pm 0.40$\\
$a_1(h2\pi^0)$& $15.94 \pm 1.73 \pm 1.7$ & $12.65 \pm 2.31 \pm 0.41$\\
electron & $14.98 \pm 2.18 \pm 0,82$& $16.96 \pm 2.92 \pm 0.15$\\
muon & $14.45 \pm 2.13 \pm 1.06$  & $12.05 \pm 2.78 \pm 0.24 $\\
acol. & $13.34 \pm 3.83 \pm 1.8$ & $19.41 \pm 5.02 \pm 0.24$\\
combi. & $14.61 \pm 0.53 \pm 0.37$ & $15.52 \pm 0.71 \pm 0.09$\\
\hline\hline
\end{tabular}
}
\end{center}
\end{table}

\begin{table}[th]
\caption[]{\label{T2}
Summary on $A_\tau$ and $A_e$ measurements 
at LEP. The first error is statistical and the second is systematic
(from Ref. \cite{Alemanytp}).
}
\renewcommand{\arraystretch}{1.3}
\begin{center}
{\small
\begin{tabular}{|l|l|l|}
\hline\hline
Exp. & $A_\tau(\%)$ & $A_e(\%)$\\
\hline
ALEPH & $14.52 \pm 0.55 \pm 0.27 $ & $ 15.05 \pm 0.69 \pm 0.10$\\
DELPHI & $13.81 \pm 0.79 \pm 0.67$&
 $13.53 \pm 1.16 \pm 0.33$\\
L3 &$14.76 \pm 0.88 \pm 0.62$& $16.78 \pm 1.27 \pm 0.30$\\
$OPAL $ & $13.4 \pm 0.9 \pm 1$ & $12.9 \pm 1.4 \pm 0.5$\\
\hline\hline
\end{tabular}
}
\end{center}
\end{table}

The conceptually simplest method to determine $P_\tau$ is based on the
decay rate into a single pion Eq.(\ref{sp}).
However, 
despite their large event rates the LEP experiments are still limited 
by the statistical error.  Therefore it is important to use the maximum
number of decay channels and exploit the full multidimensional 
distribution of the two \cite{Tsai,KW,Rouge} 
and three \cite{KW,KM,Da} pion decay mode.  The
importance of exploiting as many channels as possible becomes evident
from Table \ref{T1}, the improvement from combining the four
LEP experiments is shown in Table \ref{T2}.  On the one had this result
can be used for a test of universality:
\be
\frac {A_e}{A_\tau} = 1.03 \pm 0.07,
\ee
on the other hand $A_e$ and $A_\tau$ can be combined to
\be
A_l = (14.52 \pm 0.34 ) \% .
\ee
Converted to a measurement of the effective weak mixing angle
\be
\sin^2 \theta_W = 0.23176 \pm 0.00043,
\ee
the result competes well with the measurement of SLD with
longitudinally polarized beams \cite{Baird}
\be
\sin^2 \theta_W = 0.23110 \pm 0.00029,
\ee
and is evidently an important ingredient in the combined LEP and
SLD Standard Model fit which gives \cite{Karlen}
\be
\sin^2 \theta_W = 0.23155 \pm 0.00019.
\ee
The result from $\tau$ polarization alone is about a factor five
more accurate than anticipated in the original plans for physics at LEP where
an accuracy 
$\delta \sin^2 \theta_W \le 0.002$
was considered a reasonable goal \cite{Alt}.

\subsection{\label {Ss2.3} Electric and Magnetic Dipole Moments}

As a member of the third family with a mass drastically larger
than the one of the electron or muon the tau lepton is ideally
suited for speculations about anomalous couplings 
to the photon as well as the $Z$ boson. The effects
of the electric or anomalous magnetic moments or their weak
analogues increase with the energy, hence $Z$ decays 
are particularly suited for these investigations. Constraints
on the weak couplings are derived from the decay rate or from 
$CP$ violating correlations in the decay $Z \to \tau^+ \tau^-$
\cite{Bernreuther}. By searching for an excess of hard noncollinear photons in
the radiative final state $\tau^+ \tau^- \gamma$ of the 
$Z$ boson decays
give also access to anomalous electromagnetic couplings.
\begin{table*}[tbh]
\caption[]{\label{T3}
Limits on the electromagnetic and weak dipole moments 
of $\tau$ (from Ref. \cite{Garcia})}
\begin{center}
\begin{tabular}{|c|c|c|}
\hline
& $\gamma\tau\bar\tau$ & $Z\tau\bar\tau$\\
\hline
$a_\tau$ & $-0.05<a_\tau<0.06$ &
$
\begin{array}{l}
\Re = (0.0 \pm 1.6 \pm2.3) 10^{-3}\\
\Im = (-1.0\pm3.6\pm4.3) 10^{-3}
\end{array}
$\\
\hline
$\begin{array}{c}
d_\tau \\ \!\!\![10^{-16}\mbox{e~cm}]\!\!\!
\end{array}
$ 
& $-3<d_\tau<3$ & $ |d_\tau| \le 0.1$\\
\hline
\end{tabular}
\end{center}
\end{table*}

\begin{table}[th]
\caption[]{\label{T4}
Limits on anomalous dipole moments 
of electron and muon.
}
\begin{center}
{\small
\begin{tabular}{|l|c|c|}
\hline
& $e$ & $\mu$\\
\hline
$|a-a_{Th}|\,[\mu_B]$ & $\le10^{-11}$ & $\le10^{-8}$  \\
\hline
$d [e\,\mbox{cm}]$ & $(0.18\pm0.16) 10^{-26}$ & $(3.7\pm3.4)
10^{-19}$\\
\hline
\end{tabular}}
\end{center}
\end{table}
The present experimental limits as listed in Table \ref{T3}
are by many orders of magnitude less restrictive than
these from low energy precision measurements of Table \ref{T4}.

However, general considerations suggest dramatic enhancement
of anomalous couplings with the lepton mass; at least 
proportional to $m_l^2$, eventually even $m_l^4$.
In particular in the latter case studies of the tau lepton could open
the window to completely novel phenomena.

\section{ \label {S3}Inclusive Decays, Perturbative QCD, and Sum Rules}

\subsection{ \label{S3.1}
Inclusive Decays and the Strong Coupling Constant}

The total $\tau$ decay rate into hadrons has become one of the
key elements in the measurements of the strong coupling constants.
The ratio $R_\tau$ can be expressed in terms of $\Pi^{[1]}$ and 
$\Pi^{[2]}$ which arise from the tensorial decomposition of the correlator
of the V-A current
\ba
&&R_\tau = \frac {\Gamma(\tau \to \nu + {\rm hadrons})}{\Gamma (\tau \to
e \nu \bar \nu)} = 6\pi i \int \limits_{|s|=m_\tau^2} 
\frac {{\rm d}s}{m_\tau^2}
\nonumber \\ 
&& \times \left (1-s/m_\tau \right )^2
\left [ \Pi^{[2]}(s) - \frac {2}{m_\tau^2}\Pi^{[1]}(s) \right ] 
\label{32}
\ea
with 
\ba
&& i \int {\rm d}x e^{iqx} 
\langle  0 |T[ j_\mu^{V-A}(x) j_\nu^{V-A}(0)]| 0 \rangle =
\nonumber \\
&&
g_{\mu \nu} \Pi^{[1]}+q_\mu q_\nu \Pi^{[2]}.
\ea
The contour integral receives contributions from the
region of (relatively) large $|s| \sim m_\tau^2$
 only, and $\Pi^{[1]}$ as well
as $\Pi^{[2]}$ can be calculated in perturbative QCD. Their
absorptive parts are available up to order $\alpha^{3}_{s}$.
As a consequence of the present ignorance of $c_4$, the coefficient of
the $\alpha^{4}_{s}$ term, and the large value of $\alpha_{s}$ at
the scale $m_{\tau}$ different -- at present equivalent -- prescriptions
for the evaluation lead to significantly different results for
$\alpha_{s}$ at $m_{\tau}$ and correspondingly at $M_{z}$ 
\cite{Menke}. Once ''educated guesses'' 
for $c_4$, based on dominant terms in the
perturbative series are implemented, the spread in $\alpha_{s}$
decreases significantly \cite{Maxwell}.

Recent measurements (see eg. \cite{Hoecker,Menke} and references
therein) of the differential rate
${\rm d}\Gamma/{\rm d}s$ 
where $s$ denotes the mass of the hadronic system have led to a
large number of consistency tests, strengthening the confidence
in this precision measurement of $\alpha_{s}$.  Moments 
$R_\tau^{kl}$ are
defined by introducing additional weight functions 
$(1-s/m_\tau^2)^k(s/m_\tau)^l$ which
can be measured \cite{Menke,Hoecker}
and at the same time calculated in pQCD \cite{whoelse}.

The separation of vector and axial contributions 
to the spectral function
is fairly simple for
final states with pions, $\eta$ and $\eta\prime$ only and thus of
well-defined $G$ parity.
However,
additional theoretical input is needed
for the $K\bar K \pi$ 
channel which has contributions from both $J^P=1^+$ and $1^-$. 
In fact, recent experimental results \cite{Chentp}
seem to confirm theoretical predictions \cite{FKM,KMW} of axial 
dominance.

\begin{figure*}
  \begin{center}
    \leavevmode
    \epsfxsize=8.cm
\epsffile[10 50 580 620]{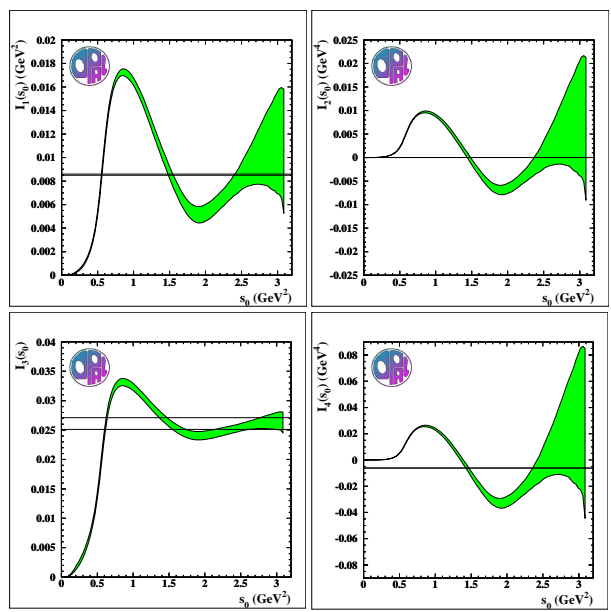}
    \hfill
    \parbox{14.cm}{
    \caption[]{\label{F6}\sloppy
Experimental results for four different sum rules, 
Eqs.(\ref{35}-\ref{38}). (From \cite{Menke}, see also \cite{Hoecker}) }}
  \end{center}
\end{figure*}

The analysis of moments
of $R_\tau^V(s)$ and $R_\tau^A(s)$ 
separately leads to an improved determination of
the vacuum condensates 
 \cite{Hoecker,Menke}, 
the weighted integrals of the difference 
\be
I_i(s_0) =\frac {1}{4\pi^2} \int \limits_{s_min}^{s_0}
{\rm d} s f_i(s) \left ( R_V(s)-R_A(s) \right )
\ee
are
particularly sensitive to nonperturbative aspects.  Experimental
results are shown in Fig.\ref{F6}
for four choices, corresponding
to four different sum rules
\ba
&&f_1(s) = 1 \Rightarrow  
\nonumber \\
&&I_1 = f_\pi^2~\left ( {\rm first~Weinberg}
\right )
\label{35}
\\
&&f_1(s) = s\Rightarrow  
\nonumber \\
&&I_2 = 0~\left ( {\rm second~Weinberg} \right )
\label{36}
\\
&&f_3(s) = \frac {1}{s} \Rightarrow 
\nonumber \\
&&I_3 = f_\pi^2 \frac {r_\pi^2}{3} - F_A,~\left ( {\rm Das, Matur, Okubo}
\right )
\label{37}
\\
&&f_4(s) = s\ln \frac {s}{\lambda^2} \Rightarrow
\nonumber \\  
&& I_4 = -\frac {4\pi}{3\alpha} f_\pi^2 \left (
m_{\pi^+}^2-m_{\pi^0}^2 \right )
\label{38}
\ea
and compared to the respective theoretical prediction.

\begin{figure*}
  \begin{center}
    \leavevmode
    \epsfxsize=8.cm
\epsffile[90 275 463 579]{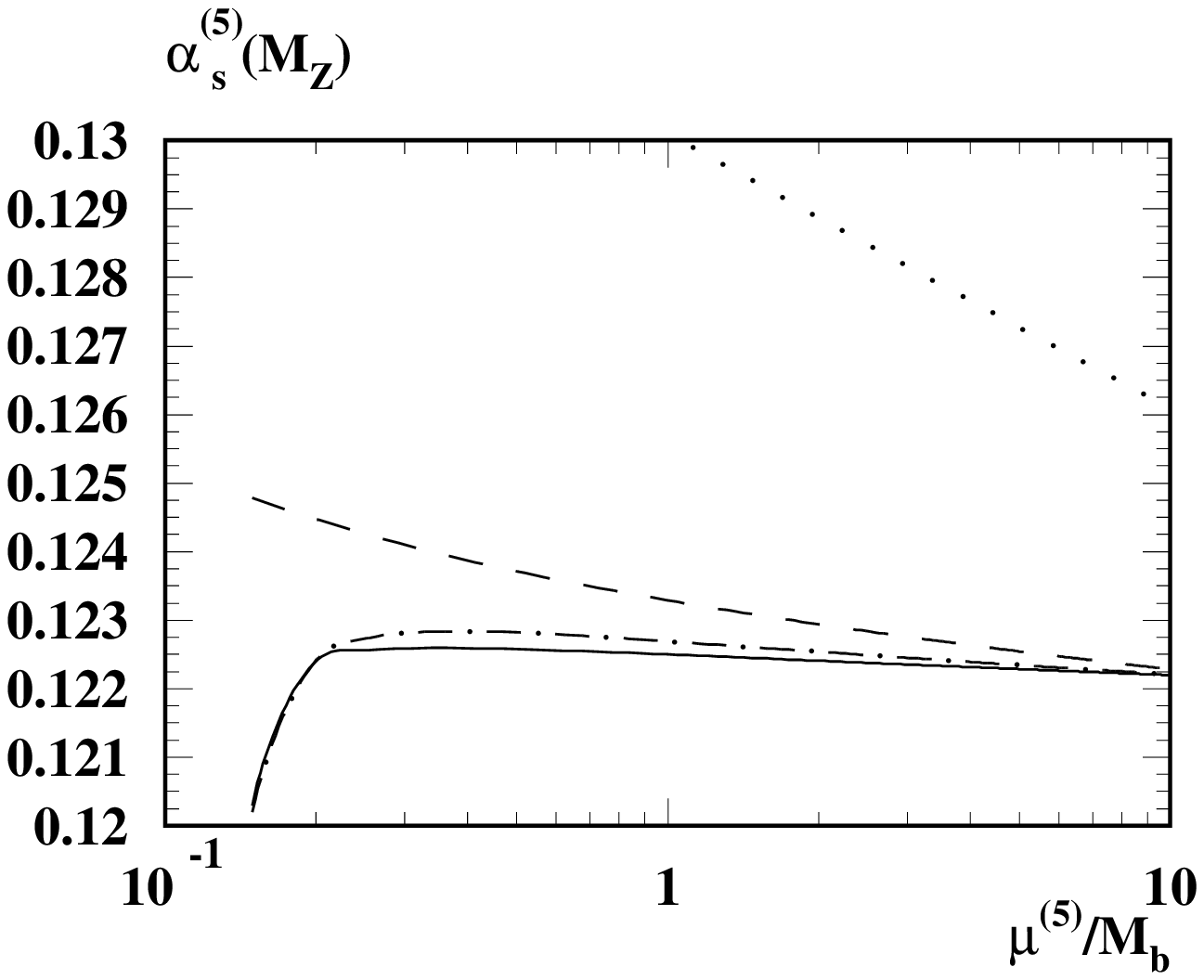}
    \hfill
    \parbox{14.cm}{
    \caption[]{\label{F7}\sloppy
$\mu^{(5)}$ dependence of $\alpha_s^{(5)}(M_z)$ calculated from 
$\alpha_s^{(4)}(m_\tau)=0.36$ and $m_b=4.7$ GeV 
using decoupling relations
at one (dotted), two
    (dashed), three (dot-dashed) and four (solid) loop
(from \cite{CKS}). 
     }}
  \end{center}
\end{figure*}

The evolution of $a=\alpha_s/\pi$ from the low scale 
$m_{\tau}$ up to $M_{Z}$ is essential for any meaningful comparison of results
obtained at vastly different energies.  Three ingredients are
crucial for any renormalization group analysis, the $\beta$ function
\be
\frac {{\rm d}\ln a}{{\rm d}\ln \mu^2} = 
-a \left (\beta_0 + \beta_1 a + \beta_2 a^2+ \beta_3 a^3 \right )
\ee
the anomalous mass dimension $\gamma$
\be
\frac {{\rm d}\ln \bar m}{{\rm d}\ln \mu^2} = 
-a \left (\gamma_0 + \gamma_1 a + \gamma_2 a^2+ \gamma_3 a^3 \right )
\ee
and the relation between the couplings $a_{n}$ and $a_{n+1}$
which are valid for the theories with n and n+1 effective quark
species respectively
\be
a_{n+1}(m_{n+1}^2) = a_n(m_{n+1}^2) \left (1
+c_2 a^2+c_3 a^3 \right ).
\ee
Important progress has been achieved during the past two years
in the calculation of all three relations: the four-loop term
of the $\beta$ function $\beta_{3}$ has been evaluated with
considerable effort and heavy machinery of algebraic programs
\cite{Verm}, similarly the corresponding four-loop term $\gamma_{3}$
of the anomalous mass dimension \cite{Chet} and, last not least, the
coefficient $c_{3}$ of the matching relation \cite{CKS}.  
Fig.\ref{F7} demonstrates
as a characteristic example the dependence of 
$\alpha_s^{(5)}(M_Z)$
on the choice of the matching point $\mu^{5}$ for the transition 
between QCD with four and five flavors respectively.  
Phenomenological studies, based on \cite{Verm,Chet,CKS} 
can also be found in \cite{RPS}.

\subsection{Cabibbo Suppressed Decays and the Strange Quark Mass}

Tau decays into final states with strangeness  proceed at the
fundamental level through the transition of the virtual $W$ to $s$ 
plus $\bar u$ quarks and are thus affected by the relatively large
strange quark mass.  After separating off the suppression through
the mixing angle $\sin ^2 \theta_C$ 
the remaining reduction of the rate in comparison
to the light quark $d$ plus $\bar u$ configuration can be attributed
to strange quark mass $m_s$.  Two important consequences are
related to $m_s \ne 0$: the divergence of both vector and axial correlator
is different from zero:
\be
q_\mu q_\nu \Pi^{VA}_{\mu\nu} = m_s^2 \Pi^{S,P}+m_s \left (
\langle \bar s s \rangle \pm \langle \bar u u \rangle \right ),
\label{42}
\ee
where $\Pi^{S,P}$ is related to the scalar correlator, which is available
\cite {askkostya} in order $\alpha^{3}_{s}$ up to the constant term.  
Eq.(\ref{42})  leads in particular 
to the production of final states with spin zero
and $q^2 \ne 0$.  However, also the rate for the $J=1$
configuration, or of
suitably chosen moments of the spectral function are significantly
affected.  The correlator for vector and axial
currents 
\be
g_{\mu \nu} \Pi^{[1]}+q_\mu q_{\nu} \Pi^{[2]}.
\ee
can always be expanded in $m_s^2/q^2$ and the
subsequent discussion is restricted to the leading quadratic mass
term.  The perturbative results can thus be cast in the following
form:
\ba
&&\Pi^{[1]} = {\rm massless~result}+\frac {3}{8\pi^2}
m_s^2 \left [l_q + a(...)+
\right.
\nonumber \\
&&\left. a^2(...) + a^3 \left (l_q^4+...+l_q + k_3^{(1)} \right ) \right ],
\\
&&\Pi^{[2]} = {\rm massless~result}-\frac {3m_s^2}{4\pi^2 q^2}
 \left [1 + a(...)+
\right.
\nonumber \\
&& \left.
a^2(...) + a^3 \left (l_q^3+...+l_q + k_3^{(2)} \right ) \right ]
\label{45}
\ea
with $a = \alpha_s/\pi$ and $l_q = \ln q^2/\mu^2$.
The constants $k_3^{[1,2]}$ 
are presently unknown.  While $k_3^{[1]}$
is irrelevant for the evaluation of the quadratic mass terms in 
the integral Eq.(\ref{32}), the constant $k_3^{[2]}$ 
contributes to the
result, a consequence of the additional factor $1/q^2$.
Higher moments, with their additional factor $s$ in the integrand can
be calculated in fixed order pQCD consistently up to order 
$\alpha^{3}_{s}$.

As stated above, the two independent quantities $\Pi^{[1]}$
and $\Pi^{[2]}$ which enter the
integral for $R_{\tau}$ carry independent information in the 
case $m_{s} \ne 0$.  It is therefore of interest to separately
consider spin zero and spin one contributions to $R_{\tau}$.
To obtain the spin zero part $R^{J=0}_{\tau}$ one has to
replace the expression in $\Pi^{[2]} - 2\Pi^{[1]}/m_\tau^{2}$
by $\Pi^{(0)}(s) - \Pi^{[1]}(0)/s$.  The first term $\Pi^{(0)}$
vanishes in the limit $m_{s} \to 0$ within perturbation theory, 
in the second term
the factor $\Pi^{[1]}(0)$ is clearly a nonperturbative object
which for the $d\bar u$ final states describes the pion contribution.
This term disappears for moments with $l \geq 1$.
Moments of $R^{J=0}_{\tau}$ with $l \geq 1$ are thus
a direct measure of the strange quark mass.

It is therefore of interest to calculate the $m_{s}$ dependence
of the rate \cite{CKw,Maltman}, of moments of the spectral function 
\cite{PP,CKP}, and
furthermore of the corresponding quantities for the spin separated
objects \cite{CKP}. 
\ba
&&\delta^{kl} = b_0 \frac {m_s^2(m_\tau)}{m_\tau^2}
\left ( 1 +b_1 a+b_2a^2+b_3 a^3 \right) = 
\nonumber \\
&&  \hat b_0 \frac {m_s^2(1{\rm GeV})}{m_\tau^2}
\left ( 1 +\hat b_1 a+...\right ),~~~k,l \ge 0
\ea
The second choice of the scale has also been tried \cite{CKP}
in order to slightly reduce the surprisingly large coefficients.
However, the reduction is only marginal  and the
subsequent discussion will be for $\mu = m_{\tau}$ only.
For the total rate one finds
\ba
&&\delta^{00}= -8\frac {m_s^2}{m_\tau^2} 
\left ( 1 + 5.33 a+46 a^2 + 248 a^3 +
\right.
\nonumber \\
&& \left. +248a^3+ 0.75k_3^{[2]} a^3 \right )
\nonumber \\
&&= -8\frac {m_s^2}{m_\tau^2} 
\left ( 1 +0.567+0.520 + 0.341 \pm 0.148 \right )
\nonumber \\
&&= -8\frac {m_s^2}{m_\tau^2} \left (2.4 \pm 0.5 \right )
\ea
where $a = \alpha_{s}/\pi = 0.106$ 
has been adopted.  The error estimate is based
on the assumption of geometric growth of the coefficients
$k^{[2]}_{i}$ in Eq.(\ref{45}).  A slight improvement of the
``convergence'' is observed for the so-called contour
improved prediction, which has been studied for the massless
case in \cite{Piv,whoelse,whoelse1}.
In this case the integral along the large circle in the complex
plane with $|s| = m_\tau^{2}$
is evaluated numerically, with $\alpha_{s}(s)$ calculated
numerically from the renormalization group with the $\beta$
function up to the four-loop coefficient.  In this case
one finds
\ba
&&\delta^{00} = -8 \frac {m_s^2}{m_\tau^2} \left (1.44 + 0.389
+0.349+
\right. \nonumber \\
&& \left.
0.0867 \pm 0.234 \right)
=-8\frac {m_s^2}{m_\tau^2} \left (2.26 \pm 0.32 \right)
\ea
The strong growth of the series can be traced to the
longitudinal piece of the correlator.  The apparent convergence
is more promising in the case of the spin one part alone.
For the lowest moment that can be evaluated in perturbation
theory one finds, for example
\ba
&&\delta^{01}|_{J=1} = -5 \frac {m_s^2}{m_\tau^2} 
\left (1.37+0.271+0.182 + 0.163 \right ) 
\nonumber \\
&&= -5 \frac {m_s^2}{m_\tau^2}
\left (2.0 \pm 0.2 \right ).
\ea
Coefficients of the higher moments with and without spin separation,
can be found in \cite{CKP}.

At the time of this workshop, 
the experimental analysis has been performed for the
total rate only \cite{Davier,Chen}.
The (unimportant) $m_{s}$ dependent terms of dimension $4$ and $6$
are available in order $\alpha^{2}_{s}$ and $\alpha_{s}$
respectively.  Applying these new theoretical results to the experimental
analysis of \cite{Davier,Chen} one obtains the result
\be
m_s(1{\rm GeV}) = \left (200 \pm 40_{\rm exp} \pm 30 _{\rm th} \right
)~~{\rm MeV},
\ee
which is in fair agreement with different, independent evaluations.
The theoretical error estimate should be considered for the
moment as a rough guess only. First steps towards the separation of $J=0$
and $J=1$ as well as the analysis of moments can be found in \cite{CDH}.

\section{\label {S4} Exclusive Decays}

\subsection{Form Factors  and Structure Functions}

The determination of form factors in exclusive decays is not only required
for a precise test of theoretical predictions. The
separation of vector and axial--vector amplitudes and their
respective spin zero and one contributions is mandatory for a number of improved
phenomenological studies like the determination of $\alpha_s$ 
and nonperturbative vacuum condensates based on moments of
vector and axial spectral functions separately, or the
unambiguous separation of vector contributions 
(in particular in the $KK\pi$ channels) for $\alpha_{QED}(M_Z)$.
Just like the search for CP violation (discussed below) or the measurement
of $g_V^\tau/g_A^\tau$ through parity violation in hadronic decays,
this can be performed with the help of a combined analysis of angular 
and energy distributions of the hadrons,
even without reconstruction of the $\tau$-restframe.

In the three (two) meson case, 
the most general ansatz for
the hadronic matrix element of the quark current,
$
J^{\mu}(q_{1},q_{2}(,q_{3}))\equiv \langle h_{1}(q_{1}) h_2(q_2) (h_{3}(q_3))
|V^{\mu}(0)-A^{\mu}(0)|0\rangle
$,
is characterized by four (two) complex form factors $F_i$, which 
are in general functions of
$s_{ij}\equiv (q_i+q_j)^2$
and $Q^2$ ($Q^\mu\equiv \sum_i q_i^\mu$):
\begin{eqnarray}
&&J^{\mu}(q_1,q_2)\!\!\! =\!\!\!  T^{\mu\nu}
\,(q_1-q_2)_{\nu}\,F     + Q^\mu\,F_S \, , \label{f2m} \\
&&J^{\mu}(q_1,q_2,q_3)\!\!\! =\!\!\!  T^{\mu\nu}
\left[ \,(q_1-q_3)_{\nu}\,F_1\,+
\right. 
\nonumber \\
&& \left. 
\,(q_2-q_3)_{\nu}\,F_2\,\right]
+
\,\,i\,
\epsilon^{\mu\alpha\beta\gamma}q_{1\,\alpha}q_{2\,\beta}q_{3\,\gamma}\,F_3
+ 
\nonumber \\
&&
Q^\mu\,F_4 \, .
\nonumber \label{f3m}
\end{eqnarray}
$T^{\mu\nu}=  g^{\mu \nu} - (Q^\mu Q^\nu)/Q^2$ denotes a transverse
projector.
The form factors $F_{1}$ and $F_{2}$  originate from the 
$J^P=1^+$ axial--vector hadronic current, and
$F_3$ ($F$) 
from the $J^P=1^-$ vector current; they
correspond to a hadronic system in a spin one state,
whereas $F_{4}$ ($F_S$)  are due to the $J=0$ part  of the 
axial--vector (vector) current matrix element. 
These form factors contain the full dynamics of the hadronic decay.
For a two--pion final state, $h_1 h_2 = \pi^- \pi^0$, 
$F_S \equiv 0$ in the isospin symmetry limit ($m_u = m_d$).
In the three pion case, $h_1 h_2 h_3 = \pi^- \pi^- \pi^+$ or
$\pi^0 \pi^0 \pi^-$, Bose symmetry 
implies $F_2(Q^2,s_{23},s_{13}) = F_1(Q^2,s_{13},s_{23})$;
$G$--parity conservation requires $F_3 \equiv 0$ for $m_u = m_d$,
and $F_4$ vanishes when $m_u= m_d = 0$.

In experimental analyses, the four(two) complex form factors 
in Eq.(\ref{f2m}) appear as sixteen (four) real
``structure'' functions $W_X$, which are defined from hadronic tensor
$H^{\mu\nu}=J^\mu J^\nu$ in the hadronic rest frame \cite{KM}.
The contributions of the different $F_i$ to $W_X$ are summarized
in Table \ref{tabsf}. After integration over the unobserved neutrino
direction, the differential decay rate in the hadronic rest frame
is given in the two-meson case by \cite{KM,FM1}
\ba
&&{\rm d} \Gamma(\tau^- \to 2h\nu_\tau) =
\{ \bar L_B W_B+\bar L_{SA} W_{SA} + 
\nonumber \\
&&\bar L_{SF} W_{SF} + \bar L_{SG} W_{SG} \}
\nonumber \\
&&\times \frac {G_F^2}{2m_\tau}\sin ^2 \theta_c \frac {1}{(4\pi)^3}
\frac {(m_\tau^2-Q^2)^2}{m_\tau^2} |{\bf q_1}| \frac {{\rm d}Q^2}{\sqrt{Q^2}}
\nonumber \\
&&\times \frac {{\rm d}\cos\theta}{2} 
 \frac {{\rm d}\alpha}{2\pi}  \frac {{\rm d}\cos\beta}{2}. 
\ea

The functions $\bar L_X$ depend on the angles and energies of the
hadrons  only \cite{KM,FM1}. The hadronic structure function
$W_X$ in the two meson case depend only on $Q^2$ and the form factors
$F$ and $F_S$ of the hadronic current. Similarly, the differential
decay rate in the three meson case is given by \cite{KM}:
\be
{\rm d} \Gamma = \frac {G_F^2}{4m_\tau} 
\left (
\begin{array}{c} 
\cos \theta_c^2\\
\sin \theta_c^2
\end{array} 
\right)
\sum \limits_{X}^{} \bar L_X W_X {\rm d}PS^{(4)}.
\label{e12}
\ee

The sum in Eq.(\ref{e12}) runs (in general) over $16$
hadronic structure functions $W_X$, which correspond
to $16$ density matrix elements for a hadronic system 
in a spin one and spin zero state (nine of them originate from a pure
spin one state and the remaining originate from a pure spin zero state or
from interference terms between spin zero and spin one, see
Table \ref{tabsf}). These structure functions contain the dynamics
of the hadronic decay and depend only on the form factors
$F_i$. Note that $W_A$, $W_B$ and $W_{SA}$ alone determine 
${\rm d}\Gamma/{\rm d}Q^2$ through
\ba
&&\frac {{\rm d}\Gamma(\tau \to h_1 h_2 h_3 \nu_\tau)}{{\rm d}Q^2}
\propto \frac {(m_\tau^2 - Q^2 )^2}{Q^4}
\nonumber \\
&&\times \int {\rm d}s_1 {\rm d}s_2 \left \{
(1+\frac {Q^2}{m_\tau^2})\frac {W_A+W_B}{6} 
+ \frac {W_{SA}}{2} \right \}.
\nonumber 
\ea

(Almost) all structure functions can be determined by studying angular
distributions of the hadronic system, for details see \cite{KM}.
This method allows to analyze separately the contribution 
from $J^{P}=0^+,0^-,1^+$ and $1^-$ in a model independent way 
(see Table \ref{tabsf}.)

In the differential decay distribution, 
the four (two) complex form factors 
appear as sixteen (four) real
``structure functions'' $W_X$, which are defined from the hadronic tensor
$H^{\mu\nu}= J^\mu J^{\nu *}$ in the hadronic rest frame \cite{KM}.
For the precise definitions of  $W_X$, we refer the reader 
to Ref.~\cite{KM}. The contributions of the different $F_i$ to $W_X$
are summarized in Table~\ref{tabsf}.
Almost all structure functions can be determined 
by studying angular distributions of the hadronic system,
which allows to analyze separately
the contributions from $J^P=0^+,0^-,1^+$ and $1^-$ in a model--independent
way.

\begin{table}
\caption{Structure functions for exclusive $\tau$ decays} \label{tabsf}
\vspace{-0.2cm}
$$
\begin{array}{|c|c|c|c|}
\hline 
& \multicolumn{3}{|c|}{H^\mu \longrightarrow}\\
H^{\nu *} \downarrow   & J^P = 1^+ & J^P = 1^- & J=0 \\
\hline \\[-3mm]
J^P = 1^+ & 
{\bf W_A}
& & \\
 &  W_C W_D W_E & & \\
\hline\\[-3mm]
J^P = 1^- & W_F W_G & {\bf W_B} &  \\
& W_H W_I & & \\
\hline\\[-3mm]
J = 0     & W_{SB}  W_{SC} 
& W_{SF} W_{SG} &
     {\bf W_{SA}}\\
 & W_{SD} W_{SG} & & \\
\hline \hline\\[-6mm]
& & \multicolumn{2}{|c|}{\underbrace{\hspace*{15ex}}_%
{\displaystyle h_1 h_2}}\\
\hline \\[-6mm]
& \multicolumn{3}{|c|}{\underbrace{\hspace*{30ex}}_%
{\displaystyle h_1 h_2 h_3}}\\
\hline
\end{array}
$$
\end{table}

\subsection{Chiral Dynamics}

Chiral Lagrangians allow for a prediction of the formfactors in the
limit of small momenta. This approach is particularly useful for 
final states with a low number, say up to three or four pions, which 
can still be treated as massless Goldstone bosons \cite{FWW}. 

In lowest order of Chiral Perturbation Theory  (CPT) the formfactors
are constants and the polynomial dependence of the amplitude is
essentially fixed by the Lorentz structure of the amplitude 
which in turn  is
imposed by the quantum numbers of the currents and the charge and
multiplicity of the pions. The same approach is also applicable
for final states with kaons or $\eta$ mesons, although their treatment as
massless Goldstone bosons and the assumption of small momenta  becomes
more dubious in these cases.

\begin{figure}
\caption[]{\label{F8}\sloppy
Differential decay rate for $\tau \to 2 \pi \nu_\tau$:
Predictions by CHPT at $O(p^2)$ (dashed), at $O(p^4)$ 
(dashed-dotted), at $O(p^6)$ (solid), and from a vector
meson dominance model (dotted), compared with exp. data 
from CLEO (dots with error bars) (from Ref. \cite{FU}).
}
\begin{center}
    \leavevmode
    \epsfxsize=8cm
    \epsffile[70 255 537 537]{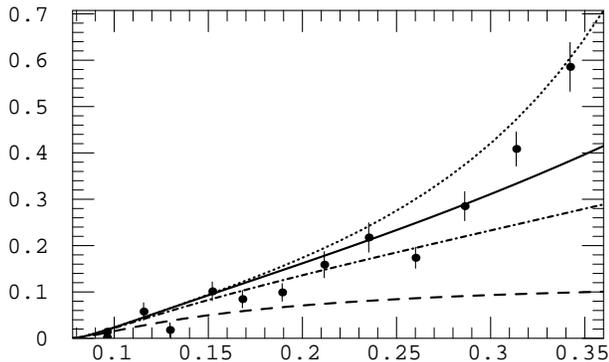}
    \hfill
  \end{center}
\end{figure}

The validity of CPT can be extended to larger momentum transfer 
by including one and even two-loop corrections \cite{FU}. For the pion
formfactor shown in Fig.\ref{F8} one observes a stronger upward curvature 
--similar to the expectation from vector dominance, where the 
lowest order chiral amplitude is simply augmented by
the $\rho$ meson 
propagator $M^2/(M^2-Q^2-i\Gamma M)$ or its more refined
variants with the proper $Q^2$ dependent width.

A more elaborated approach along the same lines has been presented
recently \cite{GuP,Gu}, which tries to combine CPT, constraints from 
analyticity
and VDM to arrive at an improved description of the data over a wide
energy range. The amplitudes, describing the decay to $\nu_\tau$
plus three pions ($\pi^-\pi^-\pi^+$ or $\pi^-\pi^0\pi^0$)
have been calculated in CPT in Born approximation \cite{FWW}and 
including one loop corrections \cite{FU}. The lowest order prediction ,
even including $a_1$ and $\rho$ mesons through vector dominance,
leads to identical predictions for both charge configurations.
Inclusion of chiral loops, however, leads to a remarkable
differences between the amplitude for the two cases, which is
in particular reflected in a difference even of the sign
of the structure function $w_D$  for small momenta \cite{FU}.

\subsection{Resonances}

The kinematical region of soft mesons, in particular pions, 
is accessible to the rigorous approach of CPT. The major fraction
of final states is, however, concentrated at intermediate 
momenta, a region dominated by a multitude of resonances.
For two pions the amplitude is well described by $\rho$ (plus $\rho
'$, etc.) dominance, with one out of many successful parameterizations
given for example in \cite{KS}. In this type of model the
normalization of the amplitude is deduced from the low energy limit,
the structure of the amplitude from the masses and widths of the
resonances. The generalization of this approach for the three pion 
case \cite{FWW,KS,Pichslac} describes successfully angular distributions,
energy spectra and distributions in the Dalitz plot, which are
conventionally encoded in the structure functions 
(see e.g. Ref.\cite{opal}).

Final states with strangeness lead to significantly more involved
substructure of the resonances. The pseudoscalar resonance in 
the configuration $\bar K^0 \pi^- \pi^0$ can have subresonances
in the $(\bar K^0 \pi^-) = \bar K^{*-}$, in the 
$(\bar K^0 \pi^0) = \bar K^{*0}$ or the $\pi^- \pi^0 = \rho^-$
system. With $G$ parity being irrelevant in this mode,
both vector and axial vector current contribute, which allows
for a nonvanishing formfactor $F_4$ (see Eq.(\ref{f2m})).
Furthermore, with $m_s \ne 0$, the currents are no longer conserved,
which allows for a small $J^P=0^-$ amplitude $F_3 \ne 0$.
The formfactors are again conveniently calculated in the 
chiral limit, the predicted total rate and the mass distributions 
depend crucially on the parameters of the resonances, their masses
and widths which are often poorly determined and must
be treated as free parameters. A typical example can be found 
in \cite{FKM,KMW}. In these papers it is shown that the branching
ratio for the decay into $K^- \pi^- \pi^+$ decreases from the 
originally predicted \cite{FM} $0.77\%$ to $0.45\%$ if both the widths
of $K_1(1400)$ and $K_1(1270)$ are increased from the PDG values
of $174$ MeV and $90$ MeV respectively to $250$ MeV.
This lower result is in acceptable agreement with the 
current world average \cite{Heltsey} of $0.304 \pm 0.074$, in
particular since no refined fit of the input parameters on the basis
of differential distributions has been performed.

\subsection{Isospin and CVC}

Important constraints on relations 
among amplitudes and decay rates  into seemingly different channels
can be derived using isospin symmetry (or, to lesser extent $SU(3)$
symmetry). In addition, one may use isospin and the hypothesis
of a conserved vector current (CVC) to relate rates 
for hadrons produced in $\tau$ decays and in electron positron annihilation.
The technique of partitions and symmetry classes, originally
developed for proton-antiproton annihilation at low energies
\cite{Pais} has been extensively applied to $\tau$ decays
\cite{Gilman,Sobie,Rouge}. The two partial rates into four pions 
$\Gamma(--+0)$ and $\Gamma(-000)$ are determined 
by the two channels for electron-positron annihilation 
into four pions and vice versa, for six pions on the other
hand only upper and lower limits can be derived. By the same kind 
of reasoning also final states with kaons can be interrelated.
For example \cite{Rouge1}
\ba
\Gamma(K^- \pi^+ \pi^-) &=& \frac {1}{2} \Gamma(\bar K^0 \pi^- \pi^0)
+ 2 \Gamma(k^- \pi^0 \pi^0)
\nonumber \\ 
2 \Gamma(K^0 K^- \pi^0 ) &\le& \Gamma(K^0 \bar K^0 \pi^-)
= \Gamma(K^+ K^- \pi^-),
\nonumber 
\ea
nicely consistent with recent experimental results \cite{Chentp}.

Up to this point only relations among decay rates have been
considered. In addition one may, in some cases, also 
exploit relations among amplitudes for different charge conjugations.
Defining, for example, for the three pion state
\be
\langle pi_1^0 \pi_2^0 \pi^- | J^{\mu} | 0 \rangle
=J^{\mu}(p_1,p_2,p^-) + J^{\mu}(p_2,p_1,p^-),
\ee
one predicts
\ba
&&\langle \pi_1^- \pi_2^- \pi^+ | J^{\mu} | 0 \rangle
=J^{\mu}(p^+,p_1,p_2)+
\\
&&J^\mu(p^+,p_2,p_1)+
J^{\mu}(p_1,p^+,p_2)+J^{\mu}(p_2,p^+,p_1).
\nonumber 
\ea
Only in the special case that the function 
$J$ is antisymmetric with
respect to the second and third variable the predictions
for the amplitude of $(0,0,-)$ and $-,-,+$ configurations 
coincide.
In general, however, differences are possible,
in particular if there is a marked resonance in 
the isospin zero configuration.

Similarly, one obtains relations between amplitude of the four-pion
final states. Given
\be
\langle \pi^+ \pi^- \pi_1^0 \pi_2^0 | J^{\mu} | 0 \rangle = 
J^{\mu}(p^+,p^-,p_1,p_2)
\ee
one predicts for $ e^+e^- \to 4 \pi$
\ba
&&
\langle \pi^+ \pi^+ \pi^- \pi^- | J^{\mu} | 0 \rangle =
J^{\mu}(p_1^+,p_1^-,p_2^+,p_2^-) +
\nonumber \\
&&J^{\mu}(p_2^+,p_1^-,p_1^+,p_2^-)
+J^{\mu}(p_1^+,p_2^-,p_2^+,p_1^-)+
\nonumber \\
&&J^{\mu}(p_2^+,p_2^-,p_1^+,p_1^-)
\ea
and for $ \tau \to 4\pi \nu$
\ba
&&
\langle \pi^- \pi^0 \pi^0 \pi^0 | J^{\mu} | 0 \rangle = 
J^{\mu}(p^-,p_1,p_2,p_3)+
\nonumber \\
&&J^{\mu}(p^-,p_2,p_1,p_3)+J^{\mu}(p^-,p_3,p_1,p_2)
\nonumber \\
&&
\langle \pi^- \pi^- \pi^+ \pi^0 | J^{\mu} | 0 \rangle = 
J^{\mu}(p_1,p^0,p^+,p_2)+
\nonumber \\
&&J^{\mu}(p_2,p^0,p^+,p_2)
\nonumber 
\ea
which connects at the same time $\tau$ decays and electron positron 
annihilation.

Analogous results are derived for final states with  kaons.
Given the amplitude for the transitions to $\pi^-\pi^0\bar K^0$
and $\pi^0\pi^0 K^-$
\ba
&&\langle \pi^- \pi^0 \bar K^0 | J_{\mu} | 0 \rangle
=J_{\mu}^{I=1}(q_{-},q_0,q_K),
\\
&&\langle \pi^0 \pi^0 K^- | J_{\mu} | 0 \rangle
=J_{\mu}^{I=0}(q_{1},q_{2},q_K),
\ea
which leads necessarily to the two pions 
in the $I=1$ and $I=0$ configurations.
The amplitudes are odd
under $q_{-} \leftrightarrow q_0$ (even under $q_1 \leftrightarrow q_2$)
and one obtains: 
\ba
&&\langle \pi^+ \pi^- \bar K^- | J_{\mu} | 0 \rangle
=
\\
&& J_{\mu}^{I=0}(q_{+},q_{-},q_K) -\frac {1}{\sqrt{2}}
J_{\mu}^{I=1}(q_{+},q_{-},q_K).
\nonumber 
\ea

\subsection{Hadronic Vacuum Polarization from $\tau$ Decays}

The hadronic vacuum polarization is of central importance for the
evaluation and interpretation of various precision observables.
This applies in particular to the fine structure constant at high energies
and the anomalous magnetic moment of the muon. Both are strongly
influenced by low energy contributions to the famous $R$ ratio.
In the low energy region, it is dominated by the two pion 
state, with its amplitude given by pion form factor.

Originally, the decay $\tau \to \nu \pi^- \pi^0$ was
predicted from the pion form factor as measured
in electron positron annihilation. However, with the excellent
normalization in $\tau$ decays, 
$\delta \Gamma/\Gamma(\tau \to \nu \pi^- \pi^0)
=6\cdot10^{-3}$, significant improvement could be obtained
and the experimental error of the analysis of $a_\mu$
\cite{Alemany} is largely dominated by $\tau$ data
\be
a_\mu|_{\pi^+\pi^-} 10^{10}= \left \{
\begin{array}{cc}
495.86 \pm 12.46 & e^+e^-\\
500.81 \pm 6.03  & {\rm incl.} \tau
\end{array}
\right.
\ee

However, the transformation from $\tau$ to $e^+e^-$ data
\be
\frac {{\rm d} \Gamma(\tau \to \nu \pi^- \pi^0)}{{\rm d} s} 
\propto S_{ew} \sigma(e^+e^- \to \pi^+\pi^-)
\ee
is subject to some theoretical uncertainties which have to be
scrutinized. The short distance electroweak corrections,
collected in 
\be
S_{ew} = 1.0194 \approx 1 + \frac {\alpha}{\pi} \ln \frac {M_Z^2}{m_\tau^2}
\ee
are derived from the inclusive rate. Large distance corrections
are not included and could be channel dependent. For the cutoff
scale, a number of choices, for example 
$m_\tau,m_\rho$ or $m_\pi$ are available.

The theoretical uncertainty 
has been estimated  in \cite{MS2} around $0.5\%$.
However, already the difference between the qualitative
evaluation of $\Gamma(\tau \to \nu \pi^-)$ in \cite{MS2}
and the detailed investigation in \cite{DF} amounts to 
$1.1\%$. This could well be considered as a realistic estimate
of theoretical uncertainties from long distance correction 
effects as long as radiative corrections are based on 
inclusive rates only. (New preliminary 
from CMD-2 experiment in Novosibirsk \cite{Eidelman}
indicate a deviation of more than $3\%$.
These large differences would prevent the use of the $\tau$ 
data for precision measurements of the form factor.)
 In addition  there is the problem 
of separating
the vector and axial contributions in the $K \bar K \pi$
channel in $\tau$ decays which can be solved, at least in 
principle, by using the technique of structure functions
described in Sect.4.1. At this point it may be worthwhile
to emphasize an important difference between 
the hadronic contributions to $a_\mu$ and
$\alpha(M_Z)$. The former is largely dominated by
 very low energy data, the latter receives contributions
from the full energy region up to $M_Z$, with a weight $dQ^2/Q^2$.
This has prompted theory driven analyses 
\cite{Hoecker2,K2,Hoecker3,Erber,Groote} which employ
the $\tau$ and $e^+e^-$ data at low energies, perturbative
QCD for energies above $1.8$ GeV and away from the charm and bottom
thresholds. An important ingredient in this analysis are theoretical
results for the $R$ ratio in order $\alpha_s^2$ and $\alpha_s^3$
including quark mass effects \cite{Ch1,Ch2,H1}.

The relative size of various contributions to 
$\Delta \alpha(M_Z)|_{\rm had}$ is listed in Table \ref{T6},
a compilation of recent results on 
$\Delta \alpha(M_z)|_{\rm had}$ from different authors is shown in
Table \ref{T7}.

\begin{table*}[th]
\caption[]{\label{T6}
Contributions to $\Delta\alpha^{(5)}_{\rm had}(M_Z^2)$
from different energy regions. From Ref. \cite{K2}.
}
\renewcommand{\arraystretch}{1.3}
\begin{center}
{\small
\begin{tabular}{|l|l|r|}
\hline\hline
Input & energy region & $\Delta\alpha^{(5)}\times 10^4$ \\ 
\hline
low energy data~\cite{Hoecker2} & $2m_\pi-1.8$~GeV & $   56.90 \pm     1.10$\\
narrow charmonium resonances & $J/\Psi, \Psi(2S), \Psi(3770)$ &
 $    9.24 \pm     0.74$\\
``normalized'' data & $3.7-    5.0$~GeV & $   15.67 \pm     0.34$\\
$\Upsilon$ resonances & $\Upsilon(1S)-\Upsilon(11.019)$ &
 $    1.17 \pm     0.09$\\
interpolation of $b\bar{b}$ & $11.075-11.2$ & $    0.03 \pm     0.03$\\
pQCD (and QED) & $1.8-\infty $ & $  194.45 \pm 0.96$\\
\hline
& total & $  277.45 \pm 1.68$\\ 
\hline\hline
\end{tabular}
}
\end{center}
\end{table*}

Although $\tau$ data and pQCD have lead to significant
improvements, independent confirmation from electron 
positron colliders is highly desirable. The low energy region,
being of prime importance for the muon anomalous magnetic moment,
will be exploited by VEPP-2M and DAPHNE, intermediate energies
up to $5$ GeV at BES2. Improved measurement at CESR or the
$B$-factories could further strengthen our confidence in pQCD
at the relatively high energy of $10$ GeV.

The $\phi$-factory DAPHNE and the $B$-factories plan
to operate primarily at their designed energy around $1$ GeV and $10$ GeV.
Using radiative events with {\em tagged} photon, one could 
well explore a large energy region, with $\sqrt{Q^2}$ varying from 
the two pion threshold up to $\sqrt{s}$. 

The differential cross section
\ba
&&s\frac {{\rm d} \sigma }{{\rm d}Q^2 {\rm d} \cos \theta}
= \frac {\alpha}{2\pi}\sigma(Q^2) \frac {1}{1-\cos^2 \theta_\gamma}
\nonumber \\
&&
\times \left ((1+\cos^2 \theta_\gamma ) 
\underbrace{\frac {(s-Q^2)}{s}}_{\sim 1} 
+ \underbrace {\frac {4Q^2}{s-Q^2}}_{\ll 1} \right )
\ea
is proportional to $\sigma(Q^2)$, the cross section for the 
$e^+e^-$ annihilation into corresponding hadronic state 
at lower energies. The factor $1/(1-\cos^2 \theta _\gamma)$ leads
to predominance of initial state radiation versus final state 
radiation for small photon angles.
The $Q^2$ dependence of the last factor is fairly flat in the 
region $Q^2 \ll s$. Since $\sigma(Q^2)$ is integrated with a flat
weight (again for $Q^2 \ll M_z^2$) 
\be
\Delta \alpha_{\rm had}(M_z^2) \sim \int {\rm d}Q^2 \sigma (Q^2)
\left [\frac {M_z^2}{(M_z^2-Q^2)} \right ]
\ee
the photon energy resolution is not necessary crucial for the
analysis. The detailed analysis, based on a Monte Carlo generator
\cite{Als} demonstrates that initial and final state radiation 
can be reasonably well separated, if proper cuts on pion and photon
angles are imposed.

\begin{table}[th]
\caption[]{\label{T7}
Comparison of different evaluations of $\Delta\alpha^{(5)}_{\rm had}(M_Z^2)$.
(${}^*\Delta\alpha_{\rm top}(M_Z^2)$ subtracted.) From Re.\cite{K2}
}
\renewcommand{\arraystretch}{1.3}
\begin{center}
{\small
\begin{tabular}{|l|l|}
\hline\hline
$\Delta\alpha^{(5)}_{\rm had}(M_Z^2)\times 10^4$ & Reference \\
\hline
$273.2 \pm 4.2$   & \cite{MarZep95}, Martin et al. `95 \\
$280 \pm 7$       & \cite{EidJeg95}, Eidelman et al. `95\\
$280 \pm 7$       & \cite{BurPie95}, Burkhardt et al. `95\\
$275.2 \pm  4.6$  & \cite{Swa95},    Swartz `96\\
$281.7 \pm 6.2$   & \cite{AleDavHoe97}, Alemany et al. `97\\
$278.4 \pm 2.6^*$   & \cite{Hoecker2}, Davier et al. `97\\
$277.5 \pm 1.7$   & \cite{K2}, K\"uhn et al. `98 \\
\hline\hline
\end{tabular}
}
\end{center}
\end{table}

\section{ \label{S5}Beyond the Standard Model}

\subsection{CP Violation in Hadronic $\tau$ Decays}

CP violation has been experimentally observed only in the
$K$ meson system. The effect can be explained by a nontrivial complex
phase in the CKM flavour mixing matrix. 
However, the fundamental origin of CP violation
is still unknown.
In particular the CP properties of the third fermion
family are largely unexplored.
Production and decay of $\tau$ leptons might offer a particularly clean
laboratory to study these effects.
CP violation 
which could arise in a framework outside the conventional mechanism 
could be observed \cite{KM2} in semileptonic $\tau$ decays.
The structure function
formalism Sect.4.1 allows for a systematic analysis
of possible CP violation effects in the two and three
meson channels.  Special emphasize is put on the $\Delta S=1$
transitions $\tau \to K \pi \nu_\tau$, where possible CP
violating signals from multi Higgs boson models would be signaled
by a nonvanishing difference between the structure function
$W_{SF}(\tau^- \to (K\pi)^- \nu_\tau)$ and its charge conjugate. 
Such a measurement is possible for
unpolarized single $\tau$'s without reconstruction of the 
$\tau$ rest frame and without polarized incident $e^+e^-$ beams.
This CP violation requires both non-vanishing hadronic phases
and $CP$ violating phases in the Hamiltonian. 
The hadronic phases arise from the interference of complex Breit-Wigner
propagators, whereas the CP violating phases could arise 
from an exotic charged Higgs boson. An additional independent
test of CP violation in the two meson case is possible,
but would require the knowledge of the full kinematics
and $\tau$ polarization (or correlation studies).
 
Transitions from the vacuum to pseudoscalar mesons $h_1$ and $h_2$ 
are induced through vector current and scalar currents
only. The general amplitude for the $\Delta S = 1$ decay
(where $h_1 = K$ and $h_2 = \pi$)
\be
\tau^-(l,s) \to nu(l',s')+h_1(q_1,m_1)+h_2(q_2,m_2),
\ee
thus can be written as
\ba
{\cal M} = \sin \theta_c \frac {G_F}{\sqrt{2}} \bar u(l',s')
\gamma_\alpha (1-\gamma_5) u(l,s) 
\nonumber \\
\times 
\left [
(q_1-q_2)_\beta T^{\alpha \beta}F + Q^\alpha \hat F_S \right ]
\label{20}
\ea
with 
\be
\hat F_S = F_S + 
\frac {\eta_S}{m_\tau} F_H.
\label{21}
\ee
In Eq.(\ref{20}) $s$ denotes the polarization four-vector
of the $\tau$ lepton. The complex parameter $\eta_S$
in Eq.(\ref{21}) transforms like
\be
\eta_S \to  \eta_S^*
\ee
and thus allows for the parameterization of possible $CP$ violation.

$W_{SF}$ can be measured in 
$e^+e^-$ annihilation experiments through the study of single unpolarized
$\tau$ decays, even if the $\tau$ rest frame cannot be 
reconstructed~\cite{KM2}.
This differs from earlier studies where either polarized beams and 
reconstruction
of the full kinematics \cite{tsaicp} or correlated fully reconstructed
$\tau^-$ and $\tau^+$ decays were required \cite{nelson1}.
The determination of $W_{SG}$, however, requires the knowledge of the full
$\tau$ kinematics and $\tau$  polarization \cite{KM2}
(eventually to be substituted through correlation studies)
which is possible with the help of vertex detectors.
The corresponding distributions in this latter case are 
similar to the correlations proposed in \cite{tsaicp,nelson1}.

The crucial observation is that one can measure the following
CP-violating differences:
\ba
\Delta W_{SF} &=& \frac {1}{2}\left (W_{SF}[\tau^-]-W_{SF}[\tau^+]\right),
\nonumber \\
\Delta W_{SG} &=& \frac {1}{2}\left (W_{SG}[\tau^-]-W_{SG}[\tau^+]\right).
\ea

One obtains
\be
\Delta W_{SF} = 4 \sqrt{Q^2} |{\bf q}_1| \frac {1}{m_\tau}
{\rm Im} (F F_H^*) {\rm Im}[ \eta_S].
\ee

In essence the measurement of $\Delta W_{SF}$ 
analyses the difference in the correlated energy distribution
of the mesons $h_1$ and $h_2$ from $\tau^+$ and $\tau^-$
decay in the laboratory. A nonvanishing $\Delta W_{SF}$
requires nontrivial hadronic phases (in addition to CP
violating phases $\eta_S$) in the form factors $F$ and $F_H$.
Such hadronic phases in $F(F_H)$ originate
in the $K\pi\nu_\tau$ decay mode from complex Breit-Wigner
propagators for the $K^*(K_0^*)$ resonance.

Once the $\tau$ rest frame is known and a preferred direction 
of polarization exists (eventually through the use of correlations
between $\tau^+$ and $\tau^-$ decays) one may also determine
\be
\Delta W_{SG} = 4 \sqrt{Q^2} |{\bf q}_1| \frac {1}{m_\tau}
{\rm Re} (F F_H^*) {\rm Im}[ \eta_S].
\ee

The sensitivity to CP violating effects in $\Delta W_{SF}$ 
and $\Delta W_{SG}$ can be fairly different depending on the hadronic
phases. First steps have been undertaken by the CLEO collaboration
to perform an experimental analysis along these lines \cite{Erklund}.

The structure function formalism \cite{KM}
allows also for a systematic analysis of possible CP--violation effects 
in the three meson case \cite{argonne}.
The $K\pi\pi$ and $KK\pi$ decay modes
with nonvanishing vector \underline{and} axial--vector
current are of particular importance for the detection of possible
CP violation originating from exotic intermediate vector bosons.
This would be signaled by a nonvanishing difference between
the structure functions $W_X(\tau^-)$ and $W_X(\tau^+)$
with $X\in\,\{F,G,H,I\}$. A difference in the structure functions
with $X\in\,\{SB,SC,SD,SE,SF,SG\}$ can again be induced through
a CP violating scalar exchange.
CP violation in the three pion channel has been also discussed in 
\cite{hagiwara} and in the $K\pi\pi$ and $KK\pi$ channels in \cite{koerner},
where the latter analysis is based on the ``$T-$odd'' correlations 
in \cite{KM} and the vector--meson--dominance parameterizations
in \cite{PI:89}.

\subsection{``Forbidden'' Decays}

A large variety of decay modes which are strictly
forbidden in the Standard Model with massless neutrinos
has been searched experimentally. They could violate
lepton family number, lepton number or baryon number,
and upper limits of a few  $10^{-6}$  have been reached
\cite{PDG}.

Heavy Dirac or Majorana type neutrinos, natural ingredients
in Grand Unified Theories, would in fact induce the 
corresponding ``Dirac type'' ($\tau^- \to l^-+...$)
or ``Majorana type '' ($\tau^- \to l^++...$) decays
through virtual corrections. The detailed theoretical
analysis performed for final states with one-meson 
\cite{Ilakovac1} and two-mesons \cite{Ilakovac2}
channels show that branching ratios around $10^{-6}$
are reasonable targets for a dedicated search program.

\section{Acknowledgments} 

Many thanks to Toni Pich and 
Alberto Ruitz for an enjoyable and productive workshop.
This paper would have never been completed without 
the \TeX nical help of K.~Melnikov.

I benefited greatly from instructive discussions with K.~Chetyrkin,
M.~Davier, A.~Hoecker, C.~Geweniger, A.~Pivovarov, A.~Stahl, 
M.~Steinhauser and N.~Wermes on various theoretical and experimental
aspects of $\tau$ physics.

\end{document}